\documentclass[aps,prb,floatfix,amsfonts,amssymb,preprint,longbibliography ]{revtex4}

\usepackage{graphicx}
\usepackage{dcolumn}
\usepackage{bm}
\usepackage{subfigure}
\usepackage{epsfig}
\usepackage{times}

\usepackage{amsmath}
\usepackage{amsfonts}
\usepackage{amssymb}

\usepackage[colorlinks=true,linkcolor=blue]{hyperref}%
\expandafter\ifx\csname package@font\endcsname\relax\else
 \expandafter\expandafter
 \expandafter\usepackage
 \expandafter\expandafter
 \expandafter{\csname package@font\endcsname}%
\fi
\begin{document}

\title{Skyrmion inertia in synthetic antiferromagnets }

\author{Sujit~Panigrahy}  \affiliation{Universit\'e Paris-Saclay, CNRS, Laboratoire de Physique des Solides, 91405, Orsay, France}
\author{Sougata~Mallick}  \affiliation{Universit\'e Paris-Saclay, CNRS, Laboratoire de Physique des Solides, 91405, Orsay, France}
\author{Jo\~ao~Sampaio}   \affiliation{Universit\'e Paris-Saclay, CNRS, Laboratoire de Physique des Solides, 91405, Orsay, France}
\author{Stanislas~Rohart} \email{stanislas.rohart@universite-paris-saclay.fr}
                          \affiliation{Universit\'e Paris-Saclay, CNRS, Laboratoire de Physique des Solides, 91405, Orsay, France}

\date{\today}

\begin{abstract}
    We describe the dynamics of magnetic skyrmions in synthetic antiferromagnets (SAF), with a finite interlayer coupling.
    Due to the opposite gyrovector of the skyrmions in the two SAF layers, the coupled skyrmions reach a stationary regime with a spatial separation, in a direction orthogonal to their velocity.
    As a consequence, and contrary to the ferromagnetic situation, a transient regime necessarily occurs, with a finite acceleration, related to an inertia, and that limits its time response.
    The formalism developed here, based on two coupled Thiele equations, allows a quantitative description of this phenomenon. The time constant associated to the transient regime scales inversely with the antiferromagnetic coupling constant.
    We also show that the coupling force reaches a maximal value at a finite skyrmion separation. This sets a maximum velocity limit, beyond which the the coupling force cannot stabilize the bound state.
\end{abstract}

\maketitle

%
Magnetic solitonic textures (domain walls, vortex cores, skyrmions...) can be described as particles that can be excited and moved by external forces. Therefore, similarly to real particules, the question of a mass and inertia naturally rises. It was shown by D\"oring\cite{doring1948} that a texture deformation related to the velocity can be responsible for the inertia of magnetic domain walls. In a Bloch domain wall, the inner magnetization angle and the velocity appear as conjugated variables, giving rise to inertia. Magnetic skyrmions have been attracting a large interest for the last ten years, since the first proposition of skyrmion-based spintronic applications~\cite{Fert2013}. Due to their singular topology, they obey to a specific dynamics which includes a gyrotropic deflection\cite{thiele1974,everschor2011}. The question of a possible mass has been highly debated\cite{back2020}. While several theories and experiments have shown that  topological bubbles (stabilized without any Dzyaloshinskii Moriya interaction) display an inertia\cite{moutafis2009,makhfudz2013,buttner2015,makhfudz2013}, the possibility of a mass in chiral and small skyrmions remain elusive and debated.\cite{back2020} Some studies report the absence of a mass in ferromagnetic skyrmions\cite{kravchuk2018}, while some cases observe an inertia related to a specific excitation\cite{shiino2017,wang2021} or to the vicinity of magnetic defects\cite{plavis2020}. In contrast, the recent interest for antiferromagnetic systems\cite{jungwirth2016,jungwirth2018,manchon2019} bring a new type of skyrmion, where a mass, intrinsic to the texture, naturally arises from the antiferromagnetic coupling\cite{velkov2016,kravchuk2018,kim2017a,kim2017b}.

Among the antiferromagnetic systems, synthetic antiferromagnets (SAF) are quite appealing to the development of skyrmions, since they take the most of ferromagnetic systems, where skyrmion are now routinely stabilized\cite{jiang2015,boulle2016,Moreau-Luchaire2016,Soumyanarayanan2017,hrabec2017}.  In SAF, two ferromagnetic layers are coupled antiferromagnetically through a non-magnetic spacer, via RKKY interaction, which allows a tunable antiferromagnetic coupling. Predicted to display a vanishing skyrmion gyrotropic deflection~\cite{barker2016,Zhang2016} similarly to real antiferromagnetic skyrmions\cite{barker2016}, they enable the stabilization of small skyrmions\cite{legrand2020} that can be moved, when excited by spin-orbit torques\cite{dohi2019}.

In this paper, we study the skyrmion dynamics in synthetic antiferromagnets and the consequences of their natural inertia. We derive an analytical expression of the skyrmion coupling force, including the dependence with the antiferromagnetic coupling strength and the skyrmion size.
The inertia arises from the opposite gyrovectors in the two layers, due to the antiferromagnetic alignement, which spatially separates the two skyrmions, in a direction perpendicular to the applied force. The skyrmion separation therefore appears as the variable  conjugated to the skyrmion pair velocity. The associated mass is inversely proportional to the coupling strength and decreases when the skyrmion size increases.  We describe the transient regime associated to this inertia, both for the bound state velocity and for the skyrmion separation, with transient times of the order of a nanosecond. Our approach allows investigations beyond the linear approximation used in effective models~\cite{velkov2016,kim2017a,kim2017b}, and predicts a maximum skyrmion velocity beyond which the bound state is destroyed under the action of the gyrotropic forces in each layer (the two skyrmions forming the pair behave as independent particles).

\section{Thiele equation modeling}

We consider two antiferromagnetically coupled magnetic layers, hosting N\'eel skyrmions that form skyrmion pairs or bound states. To investigate their  dynamics, we explicitly describe each skyrmion, rather than using an effective model that represent the system on a single layer with effective parameters~\cite{velkov2016,kim2017a,kim2017b}.  We use the Thiele formalism, initially developed for ferromagnetic systems, that describes the dynamics of a rigid texture $\mathbf{m}(\mathbf{r},t)=\mathbf{m}_0(\mathbf{r}-\mathbf{v}t)$ ($\mathbf{m}$ is the local and time-dependent magnetization, $\mathbf{v}$ its velocity and $\mathbf{m}_0$ is the stationary texture profile)~\cite{thiele1973,thiele1974}. In the two coupled layer system, we use two coupled Thiele equations
\begin{subequations}
    \label{eq:2Thiele} 
    \begin{eqnarray}
     \mathbf{G}\times\mathbf{v}_1-\alpha D\mathbf{v}_1 + \mathbf{F}_1 + \mathbf{F}_{2/1} &=& 0 \\
    -\mathbf{G}\times\mathbf{v}_2-\alpha D\mathbf{v}_2 + \mathbf{F}_2 + \mathbf{F}_{1/2} &=& 0.
    \end{eqnarray}
\end{subequations}
where $\mathbf{v}_1$ and $\mathbf{v}_2$ are the skyrmion velocities in each layers.
This extends the Thiele formalism to a coupled system\cite{barker2016,buttner2018} using the coupling forces $\mathbf{F}_{2/1}$ and $\mathbf{F}_{1/2}$ that account for the action of the textures on one another. Note that the action/reaction principle implies $\mathbf{F}_{2/1}=-\mathbf{F}_{1/2}$. The three other terms in the equations, similar to those in single ferromagnetic layers, correspond respectively to the gyrotropic deflection (with $\mathbf{G} = L_St \; n  \mathrm{~}\mathbf{z}$ the gyrovector,  $L_S=M_S/\gamma$ the angular momentum density, $M_S$ the spontaneous magnetization, $\gamma$ the gyromagnetic ratio and $t$ the film thickness), the energy dissipation (with $\alpha D   = \alpha L_St \; d$ and $\alpha$ the Gilbert damping parameter) and the force of an external stimulus.  Here, we consider the motion induced by a spin-orbit torque (SOT) due to a charge current $\mathbf{j}$ which flows in an adjacent layer along the $x$ direction.\cite{sampaio2013} This charge current is converted to a spin current $\theta_H\mathbf{j}\times\mathbf{y}$, via the spin Hall effect (SHE) ($\theta_H$ is the spin Hall angle), polarized along the $y$ direction \cite{miron2011,liu2012} and leads to a force
$\mathbf{F} = -\frac{\hbar}{2e}\mathbf{j}\theta_{H} \; f$ ($\hbar$  the reduced Plank constant and $e$ the elementary charge).
The three parameters $n$, $d$ and $f$ are related to the texture profile, as
\begin{subequations}
    \label{eq:Thiele parameters ndf}
    \begin{eqnarray}
 n &=& \iint \left( \frac{\partial \mathbf{m_0}}{\partial x}\times\frac{\partial \mathbf{m_0}}{\partial y} \right)\cdot\mathbf{m_0}d^2r\\
 d &=& \iint \left( \frac{\partial \mathbf{m_0}}{\partial x} \right)^2d^2r\label{eq:Thiele parameters d}\\
 f &=& \iint \left( m_0^{(x)}\frac{\partial m_0^{(z)}}{\partial x}-m_0^{(z)}\frac{\partial m_0^{(x)}}{\partial x} \right)d^2r
    \end{eqnarray}
\end{subequations}
with $m_0^{(i)}$ the components of $\mathbf{m_0}$. The skyrmion profile is essentially described by four parameters, the skyrmion radius $R$, the micromagnetic domain wall width parameter $\Delta$ (the lengthscale over which the magnetization rotates~\cite{rohart2013}), the skyrmion core polarization $p$ and the chirality $c$ ($c = \pm1$ respectively for clockwise and counterclockwise spin rotation), which leads to simple expressions for $n$, $d$ and $f$.
The number $n$ corresponds to the skyrmion topology and only depends on the skyrmion core polarization, with $n=4\pi p$.
The number $d$ is similar to an exchange integral and involves the magnetization rotation lengthscale. For $R\gg\Delta$, the skyrmion has a bubble profile~\cite{rohart2013} and $d=2\pi R/\Delta$~\cite{sampaio2013}. For $R\ll\Delta$, the integral does not go to zero due to the non-trivial skyrmion topology~\cite{belavin1975} and $d\rightarrow4\pi$ for $R\rightarrow 0$~\cite{buttner2018}. Over the full size range, $d$ can be approximated by
$d \approx4\pi[\exp\left(-\frac{R}{2\Delta}\right)+\frac{R}{2\Delta}]$
to account for the two limits\cite{berges2022,note_dissipation}.
The number $f$ is similar to a Dzyaloshinskii-Moriya interaction (DMI) exchange and involves the skyrmion chirality, with $f\approx \pi^2cR$. In equation~\ref{eq:2Thiele}, $\mathbf{G}$ and $D$ are calculated from the texture in layer 1. Since layer 2 is aligned in the opposite direction, the gyrovector is opposite ($n$ is odd in $\mathbf{m}$) but the dissipation is the same ($d$ is even in $\mathbf{m}$). Even if $f$ is even in $\mathbf{m}$, we consider the possibility of two different SOT-induced forces since, in a SAF stack, the torques can hardly be identical
(the spin current source being an adjacent layer, it can hardly be the same for both magnetic layers since, for one of them, the spin current source is in the spacer and therefore thinner than for the other one).

\section{Interlayer coupling between two skyrmions}

The coupling between the two skyrmions originates from the antiferromagnetic interaction between the two layers, with the energy  $E_\mathrm{AF}= J_\mathrm{AF}\iint (\mathbf{m}_1.\mathbf{m}_2+1)d^2r$, where $\mathbf{m}_1$ and $\mathbf{m}_2$ are the skyrmion profiles in each layer and $J_\mathrm{AF}$ is the interaction energy constant (positive to promote the antiparallel alignement). For skyrmions with opposite core magnetization, this energy is minimized when the skyrmions are aligned in the $(x,y)$ plane. When separated by $\mathbf{\delta R}$, the energy increases.

For a small separation as compared to $R$ and $\Delta$, i.e. when there is a significant overlap between the area where the magnetization rotates (skyrmion periphery),
$\mathbf{m}_2\approx-\mathbf{m}_1-\delta R\partial_u \mathbf{m}_1-\frac12\delta R^2\partial^2_u \mathbf{m}_1$
(with $\delta R$ the magnitude of the skyrmion separation and $u$ the coordinate along the separation direction) so the interaction energy can be developed as
\begin{eqnarray}
E_{AF}^{small} &\approx& \frac{J_\mathrm{AF}}{2}\delta R^2\iint \left( \frac{\partial \mathbf{m}_1}{\partial u} \right)^2d^2r
               = \frac{J_\mathrm{AF}}{2}d\mathrm{~}\delta R^2
\end{eqnarray}
where the geometrical parameter $d$ (eq.~\ref{eq:Thiele parameters d}), also  involved in the dissipation, is naturally found, since it represents an exchange energy. The coupling force derives from this energy, as
\begin{equation}\label{eq:couping force linear}
\mathbf{F}^{small}_{2/1}\approx-J_\mathrm{AF}d\mathrm{~}\mathbf{\delta R}.
\end{equation}
In this linear approach, the antiferromagnetic interaction induces a spring force between the two skyrmions with $k = J_\mathrm{AF}d$ the spring constant. Note that since $d$ increases with the skyrmion radius, the spring constant increases with the skyrmion size (in the limit of $R\gg\Delta$, $k$ varies linearly with the skyrmion size).

For a large separation as compared to $\Delta$, the overlap between skyrmion peripheries is negligible. For $R>>\Delta$, the antiferromagnetic energy essentially corresponds to the overlap between the skyrmion cores, considering rigid skyrmions as circular, homogeneously magnetized domains. The associated energy
\begin{subequations}
    \begin{eqnarray}
    E^{large}_{AF}&\approx& 2J_{AF}R^2\left[\pi-2\mathrm{~}\mathrm{acos}\left(\frac{\delta R}{2R}\right)
+\frac{\delta R}{R}\sqrt{1-\left(\frac{\delta R}{2R}\right)^2}\right]\mathrm{~for~ \delta R<2R}\\
                  &\approx&2J_{AF}R^2\pi\mathrm{~for~ \delta R>2R}
\end{eqnarray}
\end{subequations}
leads to the coupling force
\begin{subequations}
\label{eq:couping force large}
    \begin{eqnarray}
\mathbf{F}^{large}_{2/1}&\approx& -8RJ_{AF}\sqrt{1-\left(\frac{\delta R}{2R}\right)^2}\mathbf{u}\mathrm{~for~ \delta R<2R}\\
&\approx& 0\mathrm{~for~ \delta R>2R}.
\end{eqnarray}
\end{subequations}%
Contrary to the first approximation, this force strength monotonously decreases with $\delta R$. Note that for $\delta R>2R$ the overlap vanishes and the coupling energy is constant, the coupling force zero and therefore the skyrmion pair is broken.

\begin{figure}[ht]
\includegraphics[width=\columnwidth]{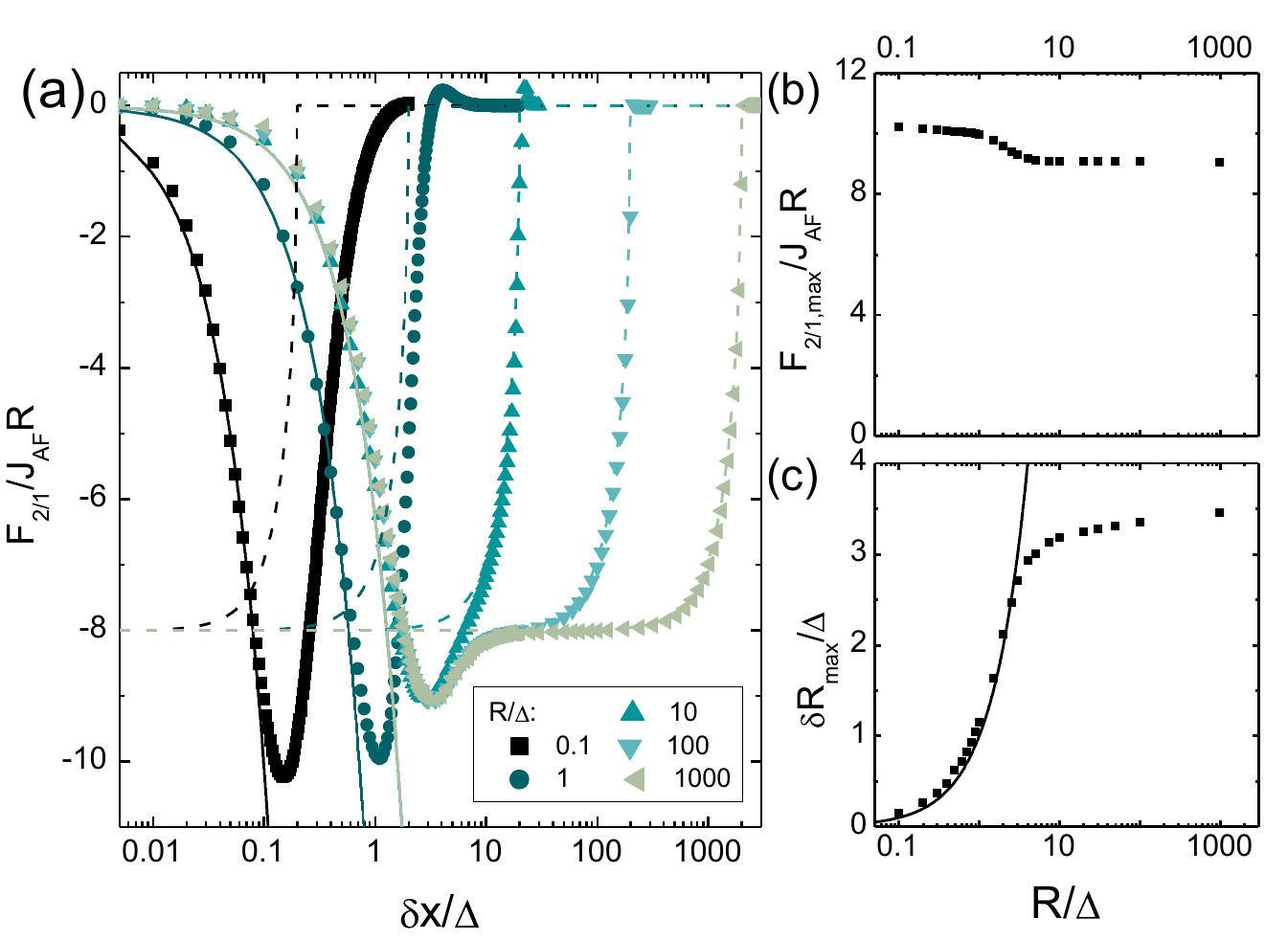}
\caption{
    (a) Skyrmion coupling force vs. skyrmion separation, for different skyrmion radius. The force is normalized to the antiferromagnetic coupling constant and the skyrmion radius. The dots corresponds to a numerical evaluation using two rigid skyrmions described by the profile obtained from the ansatz in Ref.~\onlinecite{tomasello2018}. The full lines correspond to the small skyrmion separation approximation (eq.~\ref{eq:couping force linear}), and the dashed lines correspond to the large skyrmion separation approximation (eq.~\ref{eq:couping force large}). Note that for $R/\Delta>10$ and for the smallest separations, $F_{2/1}/R$ is independent on the skyrmion size  since, in this range, $d\approx 2\pi R/\Delta$ (therefore, $F_{2/1}\approx2\pi J_{AF} R\delta R/\Delta$).
    (b) Maximum coupling force and (c) corresponding skyrmion separation vs. skyrmion size. In (c), the full line corresponds to $\delta R_\mathrm{max} = R$, valid for small skyrmions.
}
\label{fig:coupling}
\end{figure}
The two approximations are compared with a numerical evaluation (see figure~\ref{fig:coupling}a). A micromagnetic calculation of the coupling force vs. the skyrmion separation is difficult since, at a finite separation, the situation is not at equilibrium. Complex numerical approaches, such as Nudged Elastic Band methods~\cite{mills1995,dittrich2002} could be used, but are too time consuming in order to scan a broad range of parameters. Instead, under the approximation of two rigid skyrmions, the force is easily calculated by translating equilibrium profiles. In the absence of an analytical skyrmion profile, we consider the following ansatz,\cite{tomasello2018}
\begin{equation}
\tan{\frac{\theta(r)}{2}}=\frac{R}{r}\exp\left(\frac{R-r}{\Delta}\right)
\end{equation}
with $\theta$ the angle of the magnetization with the sample normal direction, and $r$ the skyrmion radial coordinate.
This formula has been shown to provide good results, whatever $R/\Delta$. Covering different skyrmion sizes ($R/\Delta$ from 0.1 to 1000), the two different regimes described above are clearly visible with a  good agreement with eqs.~\ref{eq:couping force linear} and \ref{eq:couping force large}. Only for small skyrmions ($R/\Delta\lesssim1$), the large skyrmion separation regime underestimates the coupling strength since, at such small sizes, skyrmions cannot be approximated by circular homogeneous domains. A maximum coupling force is observed [slightly larger in the numerical evaluation ($|\mathbf{F}_{2/1,\mathrm{max}|}=9$ to $10RJ_{AF}$) than  the one anticipated from the two approximations ($8RJ_{AF}$)], as plotted in Fig.~\ref{fig:coupling}b. The  skyrmion separation $\delta R_\mathrm{max}$ that corresponds to this force maximum   is of a few $\Delta$ for the largest skyrmions, and of the skyrmion radius for the smallest skyrmions (Fig.~\ref{fig:coupling}c).

\section{Steady state regime and maximum velocity}

\subsection{Skyrmion separation in the steady state}

In a steady state regime, a bound state is formed with $\mathbf{v}_1=\mathbf{v}_2=\mathbf{v}$. The solution of eq.~\ref{eq:2Thiele} is obtained by summing the two Thiele equations. In the resulting equation, the coupling force disappears and the bound state velocity is independent on the coupling strength. Additionally, the bound state displays no gyrotropic deflection in agreement with established results\cite{barker2016,Zhang2016,woo2018,hirata2019}. Therefore, the skyrmion pair moves along the total force $\mathbf{F}_{tot}=\mathbf{F}_1+\mathbf{F}_2$ direction, at a velocity $\mathbf{v} = \mathbf{F}_{tot}/2\alpha D$.

\begin{figure}
\includegraphics[width=\columnwidth]{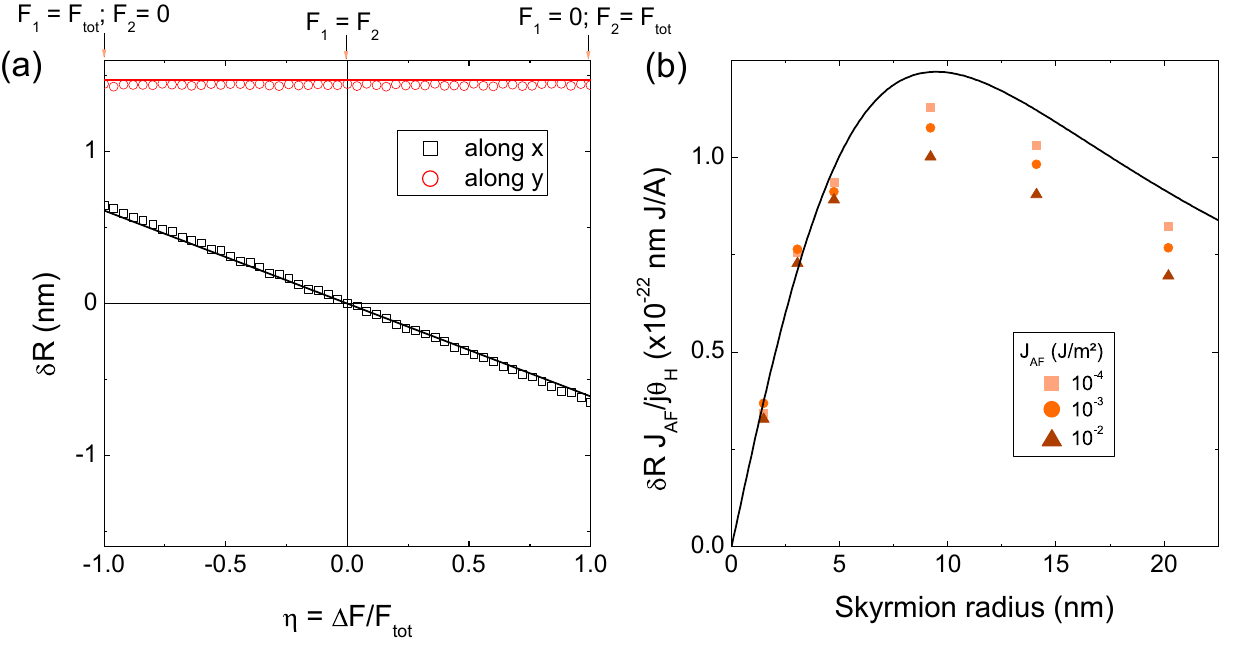}
\caption{
    (a) Micromagnetic simulation of the skyrmion separation as a function of the  SOT induced force repartition. The separation is calculated in the steady state regime for a given current $j\theta_H=3\times10^{11}$~A/m$^2$, as a function of the repartition between the two layers. The velocity (-60 m/s, not shown in the graphs) is independent of the repartition of the force. The skyrmion separation is in good agreement with eq.~\ref{eq:skyrmion separation}, and only the separation along the applied force depends on the force repartition.
    (b) Scaled skyrmion separation as a function of the skyrmion radius for different interlayer coupling constants and equal force repartition ($\eta = 0$.
    The line corresponds to the expected value from the model.
    The parameters used in the simulation are those of a CoFeB based SAF\cite{noteCFB}. In (a) the coupling constant is 0.2~mJ/m$^2$ and the skyrmion radius is 9.2~nm ($D = 2.5$~mJ/m$^2$). A rather large damping constant $\alpha = 0.3$ has been used to enhance the influence of the force difference than at more realistic values. In (b), the skyrmion radius is varied by changing the DMI from -1.4 to -2.8~mJ/$^2$.
}
\label{fig:separation}
\end{figure}

At the steady state velocity, the two skyrmions are separated, so that at the single layer level, the coupling force compensates the gyrotropic force. In the large separation regime, the coupling force strength monotonically decreases with $\delta R$, so a larger separation is unable to further compensate gyrotropic effects. Therefore, only the small separation regime is explored, with a separation limit $\delta R_\mathrm{max}$ as discussed before. If the two SOT forces are not identical, the skyrmion with the larger applied force drags the second one via the interlayer coupling, another source of skyrmion separation.
From the Thiele equations, the steady state skyrmion separation is
\begin{equation}\label{eq:skyrmion separation}
    \mathbf{\delta R} 
                      = \frac{\mathbf{G}\times\mathbf{v}           -\frac12\Delta \mathbf{F}}{k}
\end{equation}
with $\Delta \mathbf{F}=\mathbf{F}_2-\mathbf{F}_1$ the SOT force difference. The comparison with a micromagnetic simulation on CoFeB based SAF\cite{noteCFB} is shown in Fig~\ref{fig:separation} with a very good agreement. The skyrmion separation is found to be inversely proportional to the spring constant, and therefore to the antiferromagnetic coupling constant. In the specific case where $\mathbf{F}_1=\mathbf{F}_2$ ($\eta = 0$ in Fig.~\ref{fig:separation}a), $\mathbf{\delta R}$ is orthogonal to the velocity, since the skyrmion separation would only compensate the gyrotropic effects in each layer. Similarly, neglecting the gyrotropic vector, the separation would be along $\Delta \mathbf{F}$, therefore along $\mathbf{v}$, since the skyrmion separation only compensates the force difference. It is interesting to note that, in the general case, even if the skyrmion separation depends on the gyrotropic effect and the force difference, the bound state velocity is independent on the SOT force difference, as anticipated by summing the two Thiele equations.

The relative importance of the effect due to the force difference decreases at low damping constant. For a given applied total SOT force $\mathbf{F}_{tot}$, varying the force repartition between the two layers, eq.~\ref{eq:skyrmion separation} becomes
\begin{equation}\label{eq:skyrmion separation_DeltaF}
    \mathbf{\delta R} = \frac{\mathbf{G}\times\mathbf{F}_{tot}}{2k\alpha D}           -\frac{\eta \mathbf{F}_{tot}}{2k}.
\end{equation}
with $\eta$ defined to that $\mathbf{\Delta F}=\eta\mathbf{F}_{tot}$ and where the result of the steady state velocity $\mathbf{v}=\mathbf{F}_{tot}/2\alpha D$ bas been used. As a consequence, the smaller the dissipation, the closer $\mathbf{\delta R}$ from the direction perpendicular to the applied force.

The skyrmion separation is sensitive to the skyrmion size, in particular through the dependence of $k$. For a given current density, the variation is not monotonic, since $k$, $D$ and $F$ all vary with the size. This is described in Fig~\ref{fig:separation}(b), in the case of  a current equipartition ($\eta = 0$). For the smallest skyrmions, the separation varies linearly with $R$: in this regime, $d\approx4\pi$ so the increase of $\delta R$ is only due to the increase in velocity (for small skyrmions,  $\mathbf{F}_\mathrm{tot}$ increases linearly with $R$). The separation reaches a maximum as $d$ starts to vary significantly ($R>\Delta$). For the large skyrmion size $d\propto R$, as well as $k$, $D$ and $\mathbf{F}_\mathrm{tot}$, which leads to  $\delta R\propto 1/R$.

\subsection{Maximum velocity}

\begin{figure}
\includegraphics[width=0.8\columnwidth]{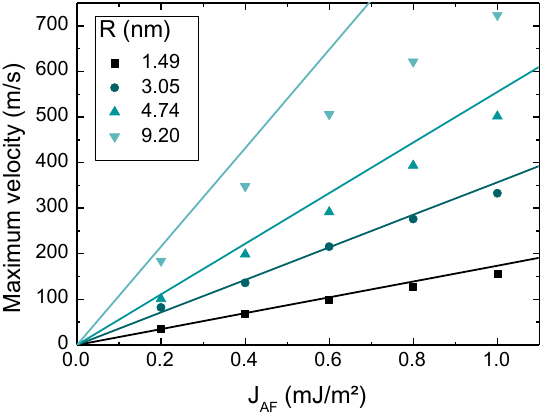}
\caption{
    Micromagnetic simulation of the maximum velocity for a skyrmion in a CoFeB based SAF\cite{noteCFB}, as a function of the antiferromagnetic coupling constant and for different skyrmion radius. The dots correspond to the simulations and the lines to the model (eq.~\ref{eq:vmax}). The SOT induced force are identical in both layers.
}
\label{fig:vmax}
\end{figure}

This preceding description, with a steady state regime, is valid as long as the coupling force can compensate the gyrotropic force. Since, the coupling force displays a maximum at finite skyrmion separation, a too large velocity implies a gyrotropic force that cannot be compensated. This leads to a skyrmion decoupling: their separation increases with time and the skyrmion dynamics is described by two independent Thiele equations. At the maximum velocity $v_\mathrm{max}$, the strengths of the gyrotropic force reaches the maximum coupling force so
\begin{equation}\label{eq:vmax}
    v_\mathrm{max} = \frac{|\mathbf{F}_{2/1,\mathrm{max}}|}{|\mathbf{G}|} \propto \frac{RJ_{AF}}{|\mathbf{G}|}
\end{equation}
where the proportionality factor varies slightly only with the skyrmion size (from 10 for the smallest skyrmions to 9 for the largest skyrmions -- see fig.~\ref{fig:coupling}b). Note that, a similar formula can be obtained using the small separation coupling force approximation, and using the maximum skyrmion separation (Fig.~\ref{fig:coupling}c) instead of the maximum force ($v_\mathrm{max} = k\delta R_\mathrm{max}/|\mathrm{G}|$). However, this approach leads to an overestimation, since at the separation maximum the coupling force exact result deviates from the linear approximation, as it can be observed in Fig.~\ref{fig:coupling}a. Eq.~\ref{eq:vmax} summarizes the physical origin of the velocity limitation in SAF skyrmions:
the maximum velocity is           proportional to the antiferromagnetic coupling constant which aligns the two skyrmions and
                     is inversely proportional to the gyrovector strength                 which pushes the two skyrmions away.
Additionally, $v_\mathrm{max}$ is proportional the skyrmion radius, which shows that pairs of larger skyrmions are better coupled (see Fig~\ref{fig:vmax}).

A situation with two different driving forces $\mathbf{F}_1$ and $\mathbf{F}_2$ ($\eta\neq0$) is another source of skyrmion separation. Therefore,  a reduction of the maximum velocity can be anticipated with
\begin{equation}
    |\mathbf{v}|_\mathrm{max} = \frac{|\mathbf{F}_{2/1,\mathrm{max}}|}{\sqrt{\mathbf{G}^2+\eta^2(\alpha D)^2}}.
\end{equation}
The maximum velocity is the highest for equal forces ($\eta = 0$) and minimum when the driving force is applied to one layer only ($|\eta| = 1$). However, the correction, proportional to $(\alpha D)^2$, is small as compared to the gyrotropic effect one, except for very large skyrmions (typically, for $R/\Delta=10$ and $\alpha = 0.1$, the maximum velocity is expected to decrease by 10~\% between the two extreme force repartition cases).

\section{Transient regime}

\begin{figure}[ht]
\includegraphics[width=0.7\columnwidth]{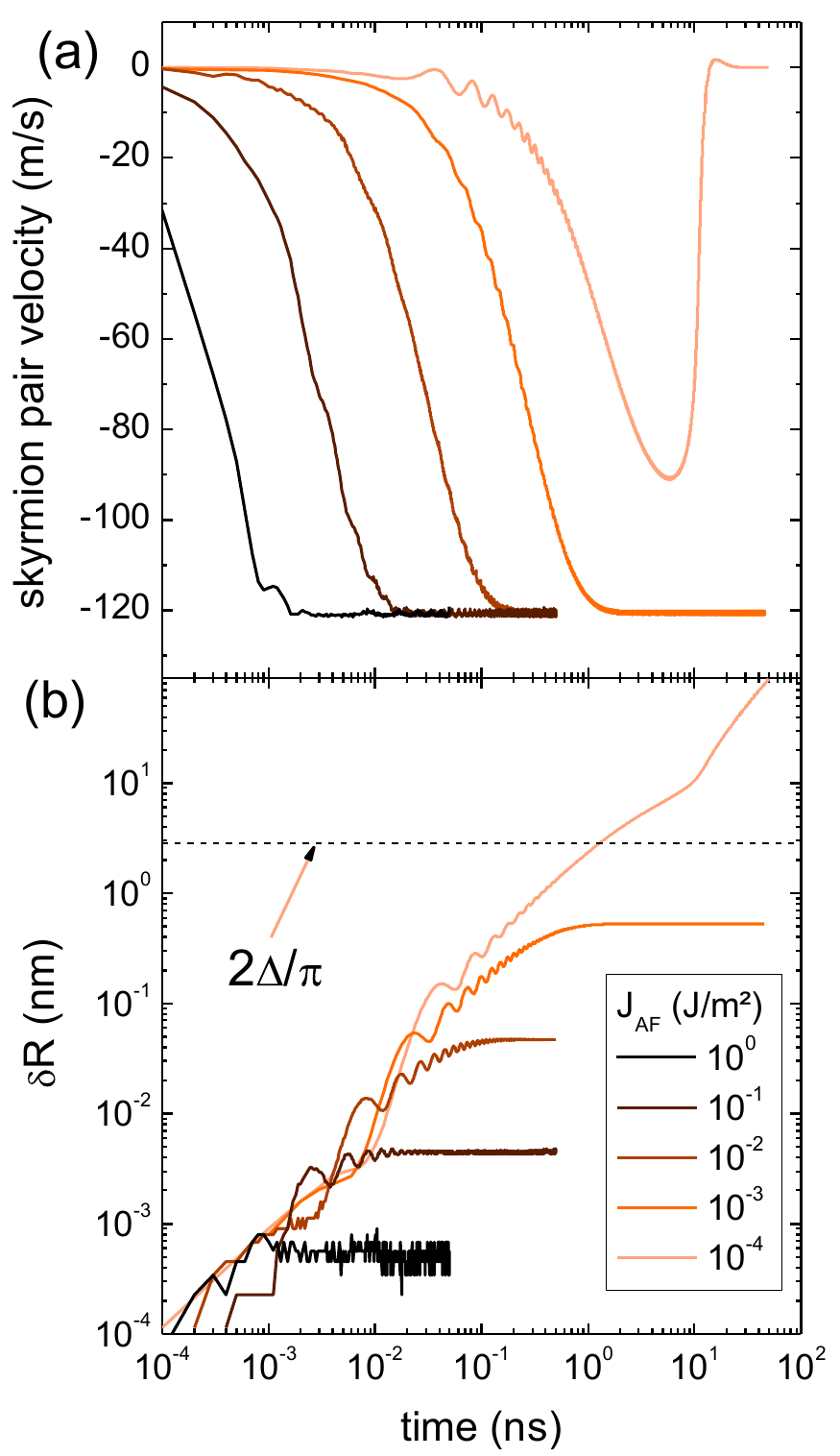}
\caption{
   Transient regime for skyrmions in a CoFeB based SAF\cite{noteCFB}. (a) Skyrmion pair velocity and (b) separation vs. time after application of the current ($j\theta_H=5\times10^9$~A/m$^2$) for different interlayer coupling constants. The skyrmion radius is 9.2~nm ($D = 2.5$~mJ/m$^2$). The dotted line corresponds to the maximum skyrmion separation according to our model. For $J_{AF} = 10^{-4}$~J/m$^2$, the skyrmion coupling is not sufficient to establish a steady state regime at this current density. Note that the velocity is negative since the skyrmion chirality is left-handed.
}
\label{fig:transient}
\end{figure}

\begin{figure}[ht]
\includegraphics[width=0.8\columnwidth]{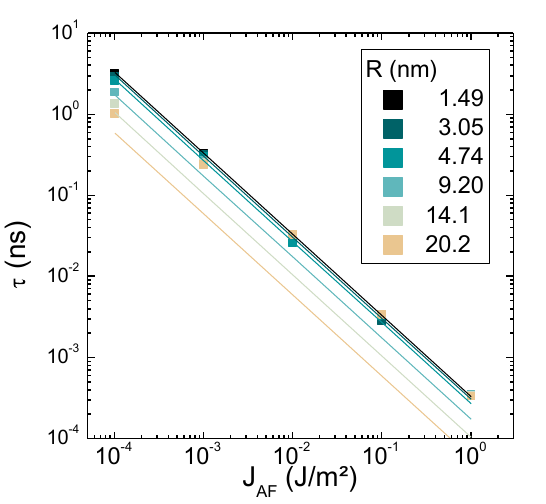}
\caption{
Transient time vs. the interlayer coupling constant for different skyrmion radius in a CoFeB based SAF\cite{noteCFB}. The points correspond to the simulation and the lines to the model (eq.~\ref{eq:tau}). The result is valid for any current densities, as long as the bound state is stable.
}
\label{fig:tau}
\end{figure}

To reach the stationary regime with spatially separated skyrmions, a transient regime is expected with independent skyrmion velocities $\mathbf{v}_1$ and $\mathbf{v}_2$, as shown in the simulations reported in Fig.~\ref{fig:transient}. Within the small separation approximation,  combining the two Thiele equations provides two independent equations for the bound state velocity $\mathbf{v}$ and the skyrmion separation $\mathbf{\delta R}$:
\begin{subequations}
    \label{eq:Thiele transient} 
    \begin{eqnarray}
    \mathbf{F}_{tot}-2\alpha D\mathbf{v}&=&\frac{\mathbf{G}^2+(\alpha D)^2}{k}\mathbf{\dot{v}}\\
    \frac{\mathbf{G}^2+(\alpha D)^2}{2k\alpha D}\mathbf{\dot{\delta R}}+\mathbf{\delta R}&=&
    \frac{\mathbf{G}\times\mathbf{F}_{tot}}{2k\alpha D}-\frac{\Delta\mathbf{F}}{2k}
    \end{eqnarray}
\end{subequations}
The first equation is similar to a single Thiele equation, but with an additional term which adds an inertia to the system, with a mass
\begin{equation}\label{eq:mass}
    M=\frac{\mathbf{G}^2+(\alpha D)^2}{k}.
\end{equation}
This inertia, which arises from the deformation of the skyrmion pair through its separation, is similar to the D\"oring mass of domain walls (in the case of the Bloch wall, the deformation corresponds to the rotation of the domain wall magnetization). It is  mostly due to the gyrotropic  effect that separates the skyrmions, and a small correction due to the energy dissipation. Note that the deformation due to the force difference, which is in fact static, does not participate to the inertia. This shows that a deformation induces inertia only if it is directly related to the velocity. The second equation describes the skyrmion separation dynamics. Here, the force difference is involved in the equation. However it has no influence on the transient regime time scale, and only affects the steady state value, as described in eq.~\ref{eq:skyrmion separation}. The two equations imply a time constant
\begin{equation}\label{eq:tau}
    \tau = \frac{M}{2\alpha D}=\frac{\mathbf{G}^2+(\alpha D)^2}{2k\alpha D}.
\end{equation}
The skyrmion pair mass and the time constant are inversely proportional to the antiferromagnetic coupling constant, in good agreement with the simulations shown in Fig.~\ref{fig:tau}. Consequently, the larger the antiferromagnetic coupling strength, the smaller the inertia and the time constant. Note that, while the dependence of the mass with the antiferromagnetic coupling has been established in Ref.~\onlinecite{velkov2016}, our approach also shows a significant dependence with the skyrmion size. Neglecting the dissipation term in the mass ($\alpha D\ll|\mathbf{G}|$), the mass is proportional to $\Delta/R$ for large skyrmions (for small skyrmions, it does not diverge and tends to $M = \mathbf{G}^2/4\pi J_AF$). The time constant is even more affected, being proportional to $(\Delta/R)^2$.

The time dependent equations also enable to understand the mechanics of the skyrmions in synthetic antiferromagnets. Combining eqs.~\ref{eq:couping force linear} and~\ref{eq:Thiele transient} in the absence of a dissipation, we obtain Hamiltonian-like equations
\begin{subequations}
    \begin{eqnarray}
        \dot{X}        &=& \frac1G\frac{\partial E}{\partial \delta Y}\\
        \dot{\delta Y} &=&-\frac1G\frac{\partial E}{\partial X}
    \end{eqnarray}
\end{subequations}
with $E$ the system total energy, $X$ the skyrmion pair position, $\delta Y$ the separation in the direction perpendicular to the direction of motion and $G = \mathbf{G}.\mathbf{z}$. These equations show that the skyrmion pair position in one direction is conjugated to the skyrmion pair separation in the other direction.

The inertia obtained from the transient regime can be compared with the skyrmion energy variation during the motion.  Within the small skyrmion separation approximation, the skyrmion pair energy in the stationary regime is
\begin{equation}
    E_{AF}= \frac12\left[\frac{\Delta\mathbf{F}^2}{4k}+\frac{\mathbf{G}^2}{k}\mathbf{v}^2\right]
\end{equation}
The first term is independent on the skyrmion velocity and only depends on the force difference. This underlines that the force difference does not participate to the skyrmion inertia. The second term, proportional to $\mathbf{v}^2$, is similar to a kinetic energy. However, the associated mass, $\mathbf{G}^2/k$, is only one part of the inertial mass since the inertia due to the dissipation does not participate to the kinetic energy. A similar fact is obtained for the domain wall mass, when comparing the D\"oring's original theory~\cite{doring1948}, that ignores damping, with more elaborated approaches.\cite{malozemoff1979magnetic}

The model is compared with simulations of synthetic antiferromagnets\cite{noteCFB} (see Figure~\ref{fig:transient}). The steady state velocity is in perfect agreement with the expected one and is independent on the interlayer coupling constant (see Figure~\ref{fig:transient}a). The transient regime observed in the simulation is more complex than what is  expected from the model, as observed in Fig.~\ref{fig:transient}. While a first order transition for the velocity and the skyrmion separation is generally observed, additional oscillations are also visible. These are induced by the magnetic precession of the regions surrounding the skyrmions when subjected to SOT.
Indeed, SOT not only pushes magnetic textures, but also acts on homogeneous regions,~\cite{miron2011,liu2012} tilting magnetic domains.\cite{liu2012b} This transition, described by the LLG equation, induces a precession of the magnetization around the new equilibrium direction. As a consequence, in the simulation, two transient regimes simultaneously occur, one for the surrounding regions and one for the skyrmion pair, and interact together. A first order law fit on the velocity or the skyrmion separation, therefore neglecting the oscillations, yields a transient time which is  close to our model expectation, in particular with a $1/J_{AF}$ dependency (see Figure~\ref{fig:tau}). However, we note that the dependency with the skyrmion size is not perfectly reproduced by the simulations, except for the smallest coupling constants. It indicates that the two transient times are of the same order of magnitude. While the transient time associated with the domain tilt is obviously independent on the skyrmion size, the one associated with the skyrmion acceleration is expected to decrease with $R$. Therefore, for the largest sizes, the transient regime is dominated by the slowest time constant, independent on the skyrmion size.

\section{Conclusion}

We have described the skyrmion dynamics in synthetic antiferromagnets, through the dynamics of two coupled rigid skyrmions. With a finite interlayer coupling, the opposite gyrotropic forces tend to spatially separate the skyrmions, which creates an inertia, inversely proportional to the antiferromagnetic coupling strength. The skyrmion coupling force displays a maximum at a finite skyrmion separation and therefore, a bound state maximum velocity is observed, beyond which the skyrmions are decoupled and behave independently. This approach can be extended to any antiferromagnetic situation that can be split into two antiferromagnetically aligned lattices. Note that in some cases (SAF with different materials and/or thicknesses, rare-earth/transition metal ferrimagnetic alloys...), the two lattices are not identical but the equations can still be solved (see appendix) with minor corrections to the inertia. The main difference between synthetic antiferromagnets and real antiferromagnets or ferrimagnets is the strength of the antiferromagnetic coupling constant. In synthetic antiferromagnets, the interfacial antiferromagnetic coupling constant is of the order of 1~mJ/m$^2$. In antiferromagnets or ferrimagnets, the volume antiferromagnetic coupling constant is larger than  $10^{8}$~J/m$^3$. This volume interaction should be integrated along the sample thickness to be used in eq.~\ref{eq:couping force linear}. For a sample thickness of a few nanometers, it leads to an interaction strength 100 times larger than in SAF and therefore to a time constant 100 times smaller, which can generally be neglected.

\appendix

\section{Inertia in unbalanced situations}

The calculations in the paper concern the inertia for a system where the two antiferromagnetically-coupled lattices are identical (except for the SOT-induced forces). For an unbalanced SAF (e.g. different layer thicknesses) or a rare-earth/transition metal ferrimagnetic alloy, the situation is different since the gyrovectors and dissipation parameters are not the same in the two lattices. With minimal approximations, the two Thiele equations can still be solved.

We consider the following pair of Thiele equations
\begin{subequations}
    \label{eq:2ThieleSAF} 
    \begin{eqnarray}
    \mathbf{G}_1\times\mathbf{v}_1-\alpha_1 D_1\mathbf{v}_1 + \mathbf{F}_1 - k\mathbf{\delta R} = 0 \\
    \mathbf{G}_2\times\mathbf{v}_2-\alpha_2 D_1\mathbf{v}_2 + \mathbf{F}_2 + k\mathbf{\delta R} = 0
    \end{eqnarray}
\end{subequations}
where the indices refer to each lattice.
Neglecting the terms in $\alpha_1 D_1-\alpha_2 D_2$, i.e. considering that the dissipation in the two lattices are close, the velocity is given by
\begin{eqnarray}
\mathbf{G}_{tot}\times\mathbf{v}-2\alpha D\mathbf{v}+\mathbf{F}_{tot}=\nonumber\\
\frac{(\Delta\mathbf{G})^2-\mathbf{G}_{tot}^2+(2\alpha D)^2}{4k}\mathbf{\dot{v}}
-\frac{\alpha D}{2k}\mathbf{G}_{tot}\times\mathbf{\dot{v}}.
\end{eqnarray}
with $\mathbf{G}_{tot}=\mathbf{G}_1+\mathbf{G}_2$, $\Delta \mathbf{G}=\mathbf{G}_1-\mathbf{G}_2$ and $\mathbf{F}_{tot}=\mathbf{F}_2+\mathbf{F}_1$. Note that in an antiferromagnetic situation, $\mathbf{G}_1$ and $\mathbf{G}_2$ are antiparallel so $|\mathbf{\Delta G}|\gg|\mathbf{G}_{tot}|$. However, contrary to the balanced antiferromagnetic situation, the two gyrovectors do not compensate and a finite deflection remains in the equation. The inertia is more complex than for the balanced situation, with an additional term. This term differs from a simple mass, as defined in classical mechanics, since it involves a cross product between the total gyrovector and the skyrmion pair acceleration. Since the gyrovector is related to the skyrmion topology, this additional inertia can be considered as a topology induced inertia. Note however that this term is small as compared to the first one and can generally be neglected so that the skyrmion pair time constant is
\begin{equation}
\tau \approx \frac{(\Delta\mathbf{G})^2-\mathbf{G}_{tot}^2+(2\alpha D)^2}{8k\alpha D}
\end{equation}
close to that of a balanced SAF.

\section{Spin transfer torque induced skyrmion motion}

Another possibility to move skyrmions in magnetic multilayers is the spin-transfer torque (STT), although it is expected to be less efficient than spin-orbit torques.\cite{sampaio2013} The calculation derived from eq.~\ref{eq:2Thiele} can also be performed with this other excitation.  The coupled coupled Thiele equations become
\begin{subequations}
    \label{eq:STT} 
    \begin{eqnarray}
     \mathbf{G}\times(\mathbf{v}_1-\mathbf{u})-D(\alpha\mathbf{v}_1-\beta\mathbf{u}) - k\mathbf{\delta R} &=& 0 \\
    -\mathbf{G}\times(\mathbf{v}_2-\mathbf{u})-D(\alpha\mathbf{v}_2-\beta\mathbf{u}) + k\mathbf{\delta R} &=& 0.
    \end{eqnarray}
\end{subequations}
where the effect of a current flowing in the magnetic layers through the magnetic  textures is represented by $\mathbf{u}$ (uniform to a velocity and proportional to the current density) and a non-adiabaticity parameter $\beta$.\cite{thiaville2005,sampaio2013} In the limit of the linear interaction regime, these equations lead to
\begin{subequations}
    \label{eq:STT-solved} 
    \begin{eqnarray}
     -2D(\alpha\mathbf{v}-\beta\mathbf{u}) &=& \frac{\mathbf{G}^2+(\alpha D)^2}{k}\mathbf{\dot{v}} \label{eq:STT-solved_v}\\
    \frac{\mathbf{G}^2+(\alpha D)^2}{2k\alpha D}\mathbf{\dot{\delta R}}+\mathbf{\delta R}&=&
    \frac1k\left(\frac\beta\alpha-1\right)\mathbf{G}\times\mathbf{u} \label{eq:STT-solved_deltaR}
    \end{eqnarray}
\end{subequations}
The first equation for the average velocity corresponds to the motion of a non-topological magnetic texture under STT, with a mass $M=\frac1k[\mathbf{G}^2+(\alpha D)^2]$. The second equation describes the dynamics of the skyrmion separation. Both equation display the same time constant $\tau=M/2\alpha D$. In the stationary regime, $\mathbf{v}=\frac\beta\alpha\mathbf{u}$, a typical result for non-topological textures (including domain walls) moved under STT\cite{thiaville2005} and $\mathbf{\delta R}=\frac1k(\frac\beta\alpha-1)\mathbf{G}\times\mathbf{u}$. This result shows that the inertial concepts derived from SOT excitation are general and not specific to a single excitation: the mass is the same and has its origin from the opposite gyrotropic effect in both layers due to the antiferromagnetic coupling.

\begin{acknowledgements}
We are grateful to Andr\'e Thiaville for enlightening discussions and a critical reading of the manuscript. This work was supported by the French National Research Agency (ANR) [under contract ANR-17-CE24-0025 (topsky) and  a public grant overseen as part of the ``Investissements d'Avenir'' program (Labex NanoSaclay, reference: ANR-10-LABX-0035), SPICY] and by an Indo-French collaborative project supported by CEFIPRA (IFC/5808-1/2017).
\end{acknowledgements}


\begin{thebibliography}{45}
\bibitem[{not({\natexlab{a}})}]{note_dissipation}
\bibinfo{note}{To account for the two limit at $R=0$
  ($d=4\pi$\cite{buttner2018}) and $R\gg\Delta$ ($d=2\pi
  R/\Delta$\cite{sampaio2013}), we represent $d$ as
  $4\pi[\exp(-R/2\Delta)+R/2\Delta)]$. In this formulation, the weight of the
  $R=0$ limit decreases exponentially with the radius, and $d$ variations are
  negligible close to $R = 0$. A comparison with micromagnetic simulations can
  be found in Ref.~\onlinecite{berges2022}.}
\end{thebibliography}
\end{document}